%% file: visas.tex
\documentclass[11pt]{article}
\usepackage[dvips]{graphicx}

\begin{document}

\begin{center}
\bf{DEPENDENCE OF SHAKE PROBABILITY ON
NUCLEAR CHARGE IN Li-, Na- AND K-LIKE IONS} 
\end{center}

\vspace{5mm}
\begin{center}
A. Kupliauskien\.{e}$^a$\footnote{Corresponding author.
Tel.: +37052612723.\\E-mail address: 
akupl@itpa.lt (A.~Kupliauskien\.{e}).},
K. Glem\v{z}a$^b$
\end{center}

\begin{center}
{\small\em
$^a$Vilnius University Institute of Theoretical Physics 
       and Astronomy,
A.Go\v{s}tauto 12, LT-01108 Vilnius, Lithuania \\
$^b$Vilnius University, Saul\.{e}tekio 9, LT-10222 Vilnius, Lithuania}
\end{center}

\vspace{10mm}
{\bf Abstract}

In sudden perturbation approximation, the  probability
of the shake-up process accompanying inner-shell ionization is
calculated for the isoelectronic sequences of Li-, Na- and
K-like ions in the ground and excited $n$p and $n$d states.
Numerical solutions of Hartree-Fock equations and hydrogen-like 
radial orbitals are used.
Very large differences between the results of both approximations
for all ions 
and strong dependences on ion charge are obtained at the beginning
of the isoelectronic sequences.

\noindent
{\em PACS}: 34.80.Hd\\
{\em Keywords}: Auger decay and inner-shell excitation or 
ionization

\input{sec1}

\include{bibl}
\include{table}

\include{figcap}

\include{figure}

\end{document}

%% file: sec1.tex
\vspace*{10mm}
\noindent
{\bf 1. Introduction}

\vspace{5mm}
A shake process accompanies inner-shell ionization of atoms and ions.
It usually effects 
photoelectron \cite{Cubaynes1992,Cubaynes1989}
and Auger electron \cite{Meyer1991,Mursu1996,Whitfield1991}
spectra.
In the case of shake-up or shake-off process, it describes the
excitation \cite{Aberg1967} or ionization
\cite{Krause} of the second electron, respectively.
The shake probability describes the relaxation of passive
electrons and is also important for the single electron ionization
cross section. 
It was obtained by Kupliauskien\.{e} \cite{K1994,K1996,K2001} that
the shake probability should be taken into account even in the case
of inner-shell ionization when spectator electrons saved their
initial states.
Then it describes the probability of spectator electrons to stay
in their initial states.

A strong increase of the relative intensity of shake-up satellites
in the inner-shell photoionization of excited Na atoms 
\cite{Cubaynes1989} was explained by the change of the 
localization
of valence electron in the  final state with respect to that
in the inital state
\cite{Felfli1992,K1994,K1996}.
Very strong dependence of the calculated inner-shell ionization cross
sections of alkaline atoms on the excited valence electron state
was noticed not only in the case of photoionization 
\cite{K1994,K1996,RK1998} but also for the ionization
of atoms by electrons \cite{Dorn1997,Nienhaus1997,RK2001}.
For neutral atoms and singly charged ions of the second and third
rows, the probability of shake process accompanying inner-shell
ionization was calculated in \cite{KG2002}, 
and very strong dependence on valence electron state was obtained.
Enormous differences were found for low excited ns and np states 
between shake probabilities that were calculated using numerical
solutions of Hartree-Fock equations (HF) and hydrogen-like 
(H-like) radial orbitals \cite{KG2002}.
A simple two-parameter interpolation function for Z-dependences
of shake-up probabilities per electron was presented by Kochur
{\em et al} \cite{Kochur2002} for atoms in the ground state.

The aim of the present work is the investigation of the dependence
of shake probability for Li-, Na- and K-like ions on the ion
charge and the state of valence electron because of the very strong
sensitivity of the inner-shell ionization cross sections of
Li-, Na- and K-like atoms on the valence electron state 
\cite{K1994,K1996,RK1998,RK2001}.
The regularities of the shake probability passing from atoms
to highly charged ions of these isoelectronic sequences are 
investigated.

The calculated shake probabilities can be used  for  searching 
 strong relaxation effects in processes dealing with
 inner-shell ionization of excited atoms and ions by
photons and electrons as well as for the Auger decay
 calculations in sudden perturbation approximation.

\vspace*{10mm}
\noindent
{\bf 2. Description of calculations}

The ionizing transition
\begin{equation}
A^{N+}(nl^{4l+2}n'l') + \alpha
\to A^{(N+1)+}(nl^{4l+1}n''l') + e^-
\end{equation}
is investigated in the present work.
Here $nl=$1s,2p,3p for Li-, Na- and K-like ions, respectively, 
$n'=n+1$ for Li- and Na-like ions, 
$n'=4$ for s- and p-electrons and $n'=3,4$ for d-electrons of K-like
ions, $l'=0,1,2$, $n''=n',n'+1$, $N$ is the charge of an ion, and
$\alpha$ stands for a particle or photon.

The shake probability can be expressed as:
\begin{equation}
P(i'\to f)= |\langle i'|f\rangle |^2
\end{equation}
where the wave function $\langle i'|$ describes the ion in the 
intermediate state following ionization, 
in which spectator electrons remain in their initial states,
with the radial orbitals of an atom, and $|f\rangle$ 
is the wave function of an ion in the final state.

The calculations of shake probabilities (2) have been performed by
using both numerical solutions of HF equations \cite{Froese}
obtained in the average-term approximation
and hydrogen-like radial orbitals
with effective nuclear charge $Z=N+1$ and $Z=N+2$ for ions
in the initial and  final states of the valence electron, respectively.

\vspace*{10mm}
\noindent
{\bf 2. Results and discussion}

The shake probabilities for single electron ionization (SPI) 
and shake-up
(SPS) transition  were calculated for Li-, Na- and K-like ions
in the configurations $nl^{4l+2}n'l'$, $l'=0,1,2$, $n'=n+1$ for
$l'=0,1$, and $n'=n+2$ for Li-, $n'=n+1$ for Na- and $n'=n,n+1$
for K-like isoelectronic sequences in the case of d-electron.
The final state of the ionized ions is $n''=n'+1,n'+2$.
The values of the shake probabilities calculated by using both HF and 
H-like radial orbitals are presented in Fig.~1--2 
for Li- and  Na-like ions, respectively,  and Table~1 for K-like ions.
The values of shake probabilities for d-electrons 
calculated with H-like radial orbitals are not
displayed in Fig.~1--2 because they are very close to those of
HF approximation for Li- and Na-like ions. 

The results of Fig.~1--2 and Table~1
demonstrate that SPI are smaller for 
d-electrons than those of s- and p-electrons.
They increase with
the decrease of the orbital quantum number from $l'=2$ down to
$l'=0$.
This tendency is independent of the charge of an ion.
The values of SPI are much smaller than those at the beginning of
all sequences investigated but rapidly increase 
up to one with growing
ion charge $N=9$, $N=8$ and $N=6$ for Li-, Na- and K-like ions,
respectively.

At the beginning of each sequence under investigation the
values of SPS are much larger than zero and decrease with
increasing ion charge.
The values of SPS  become close to zero for ion charge a little less 
than the values  of SPI which become close to one, 
i.e. 8, 7, and 5 for Li-, Na- and K-like ions, respectively.
Thus, the values of SPS decrease more rapidly than those of SPI 
increase.

The values both of SPI and SPS calculated by using H-like radial
orbitals with effective ion charge differ very much from those of HF
calculation 
not only at the beginning of isoelectronic sequences but also
for highly charged ions.
They are similar to each other only for d-electrons in the case of
Li- and Na-like ions.
In the case of Li-like ions, the differences between the results of
both approximations are less in comparison with those of
Na- and K-like ions.
In the case of K-like ions, the H-like calculations can not be
used for the evaluations of shake probabilities even for highly
charged ions (see table~1).
The results of Table~1 show that the differences between the values
calculated with HF and H-like orbitals reach several
times or even an order for both SPI and SPS  probabilities.

Calculated values of the ratios of SPS and SPI 
for Li-, Na- and K-like isoelectronic sequences are presented
in Fig.~3.
Here white points in black background stand for the experimental
ratios of shake-up and single electron  photoionization cross sections
in the case of Li \cite{Journel}, Na \cite{Cubaynes1989} and
K \cite{Cubaynes1992} atoms.
Experimental values of the ratio are chosen for large photon
energies where sudden perturbation approximation is valid.

The results from Fig.~3 show that the agreement between 
calculated ratios of shake probabilities and measured
shake-up and single electron photoionization cross sections is
good.
The values of the rations decrease with increasing ion charge.
The ratios for s- and p-electrons decrease while the number of core
electrons increases.
But the ratios for d-electrons in the case of K-like ions are
larger than those of Na- and Li-like sequences.
The ratios are larger at the beginning of the sequences,
however, their decrease is more rapid for Li-like ions than for
K-like ions.

%% file: table.tex
\noindent
Table 1.
 Shake probabilities (2) (SPI and SPS)
calculated by using HF and H-like (H) radial
orbitals with $N+1$ and $N+2$ efective nuclear charges (1)
 for K-like ions in the 3p-electron ionization 

\begin{center}
\begin{tabular}{ l l l l l l l l l l}
\hline
  &  &   &    &    &  & & & & \\
$n'l'$& $n''l'$& Method& K & Ca$^+$ & Sc$^{2+}$ & V$^{4+}$&Mn$^{6+}$ &
Co$^{8+}$& Ni$^{9+}$\\
  &  &   &    &    &  & & & & \\ 
\hline
  &  &   &    &    &  & & & & \\
4s & 4s & HF& 0.84 & 0.94 & 0.97 & 0.98 & 0.99 & 0.99 & 0.99\\
   &    & H & 0.02 & 0.20 & 0.48 & 0.75 & 0.86 & 0.91 & 0.93\\
4s & 5s & HF & 0.14 & 0.05 & 0.027 & 0.012 & 0.007 & 0.005 & 0.004\\
   &    & H & 0.67 & 0.69 & 0.41 & 0.17 & 0.085 & 0.051 & 0.042\\
4p & 4p & HF & 0.73 & 0.90 & 0.95 & 0.98 & 0.99 & 0.99 & 0.99\\
 &  & H & 0.003 & 0.25 & 0.53 & 0.78 & 0.88 & 0.92 & 0.94\\
4p & 5p & HF & 0.27 & 0.09 & 0.044 & 0.018 & 0.010 & 0.006 & 0.005\\
 &  & H & 0.69 & 0.63 & 0.37 & 0.15 & 0.077 & 0.047 & 0.038\\
3d & 3d & HF & 0.16 & 0.90 & 0.98 & 0.99 & 1.00 & 1.00 & 1.00\\
 &  & H & 0.44 & 0.75 & 0.86 & 0.94 & 0.97 & 0.98 & 0.98\\
3d & 4d & HF & 0.68 & 0.08 & 0.014 & 0.003 & 0.002 & 0.001 & 0.001\\
 &  & H & 0.54 & 0.24 & 0.12 & 0.05 & 0.026 & 0.016 & 0.013\\
4d & 4d & HF & 0.027 & 0.60 & 0.84 & 0.95 & 0.97 & 0.98 & 0.99\\
  &  & H & 0.02 & 0.39 & 0.64 & 0.84 & 0.91 & 0.94 & 0.95\\
4d & 5d & HF & 0.47 & 0.33 & 0.13 & 0.04 & 0.02 & 0.012 & 0.01\\
 &  & H & 0.70 & 0.51 & 0.29 & 0.12 & 0.06 & 0.037 & 0.030\\
  &  &   &    &    &  & & & & \\
\hline
\end{tabular}
\end{center}

%% file: figcap.tex
{\bf Figure captions:}

Fig. 1. Shake probability vs nuclear charge for Li 
isoelectronic sequence in the case of the transition
$nl^{4l+2}(n+1)l' \to nl^{4l+1}n'l'$.
Filled circles are for single electron ionization (SPI), 
diamonds are for shake-up process (SPS),
solid line stands for $l'=0$, dashed line is for $l'=1$ and dotted
line is for $l'=2$, respectively.
Triangles-dot up show single electron and triangles-dot down indicate
shake-up probabilities calculated using hydrogen-like radial orbitals
with effective nuclear charge $Z=N+1$ and  $Z=N+2$ 
(1) for
the initial and final states, respectively, and  dashed-dot and
dashed-dot-dot lines are for $l'=0$ and $l'=1$ electrons, respectively.

Fig. 2. Shake probability vs nuclear charge for 
Na  isoelectronic sequence.
Details are as in Fig.~1.

Fig. 3. The ratios of the probability of shake-up to 
that of single electron transition for Li- (filled circles),
 Na- (squares) and K-like (diamonds) isoelectronic sequences.
Solid, dashed and dotted lines are 
for s, p and d electrons, respectively.
The white dots in black background are experimental values of the
ratios between shake-up and single electron photoionization
cross sections for Li \cite{Journel}, Na \cite{Cubaynes1989} and
K \cite{Cubaynes1992} atoms.

%% file: figure.tex
Figure~1
\vspace{30mm}

\begin{figure}[h!]
\begin{center}
\includegraphics*[height=10cm,width=12cm]{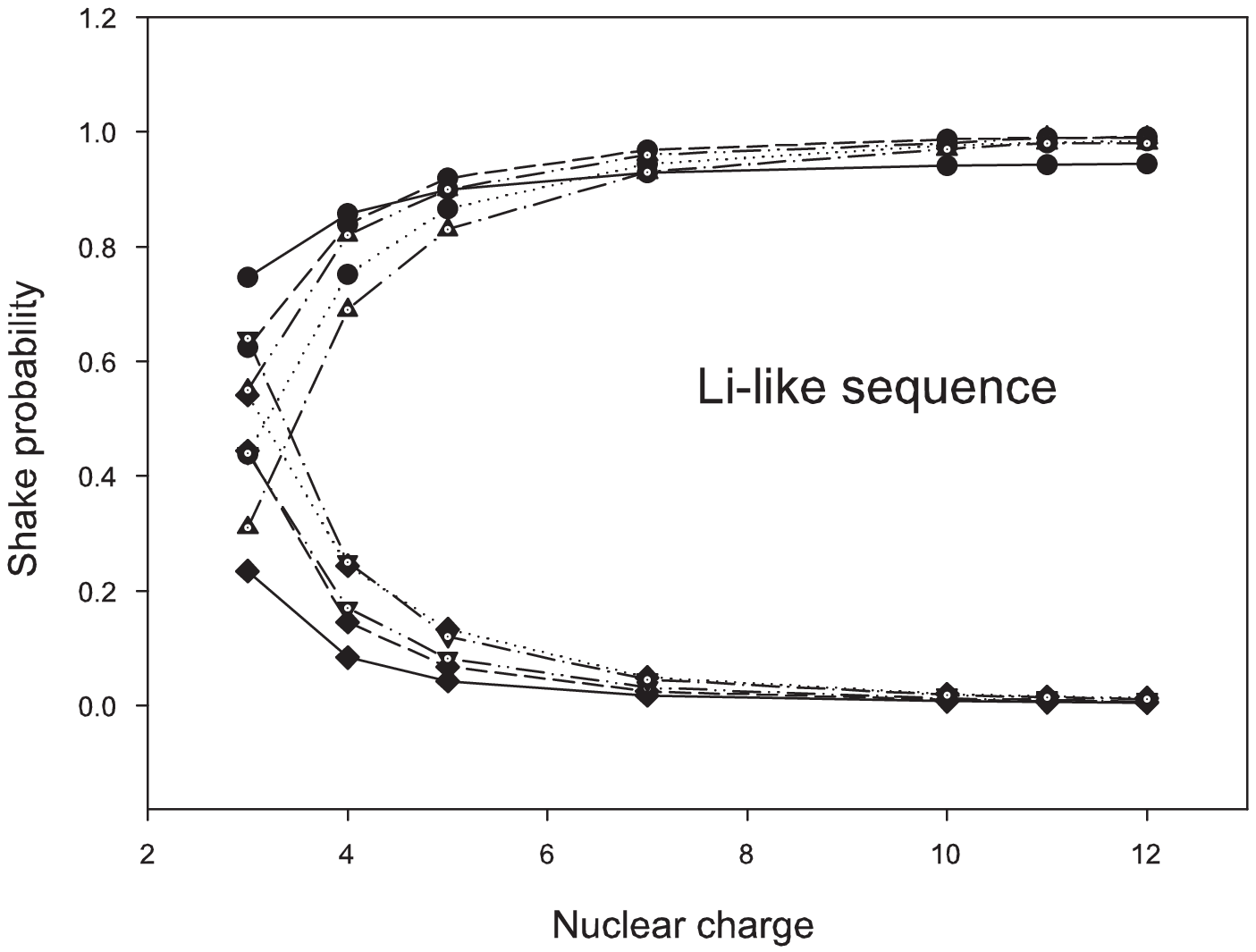}
\end{center}
\end{figure}

\clearpage
Figure~2
\vspace{30mm}

\begin{figure}[h!]
\begin{center}
\includegraphics*[height=10cm,width=12cm]{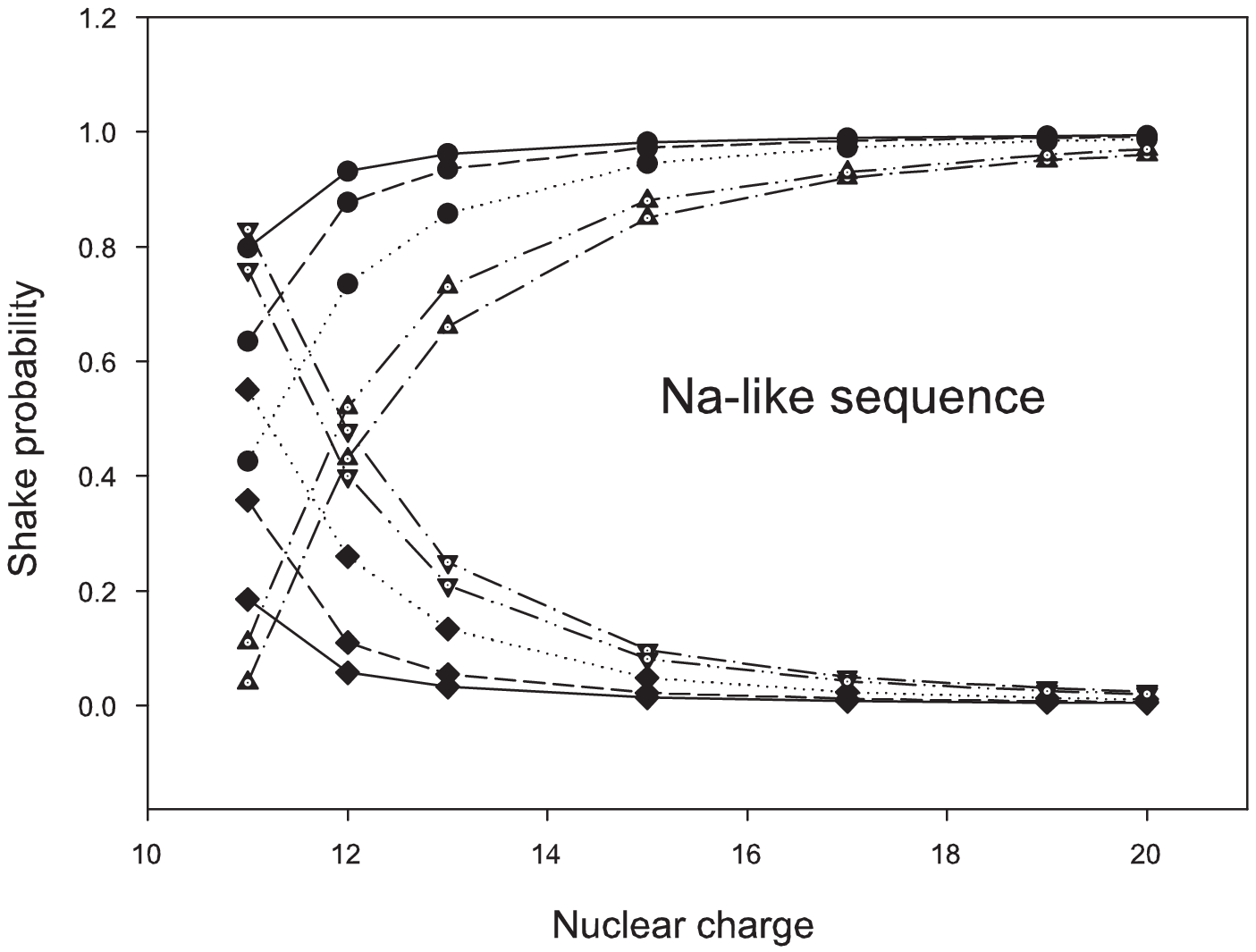}
\end{center}
\end{figure}

\clearpage

Figure~3
\vspace{30mm}

\begin{figure}[h!]
\begin{center}
\includegraphics*[height=10cm,width=12cm]{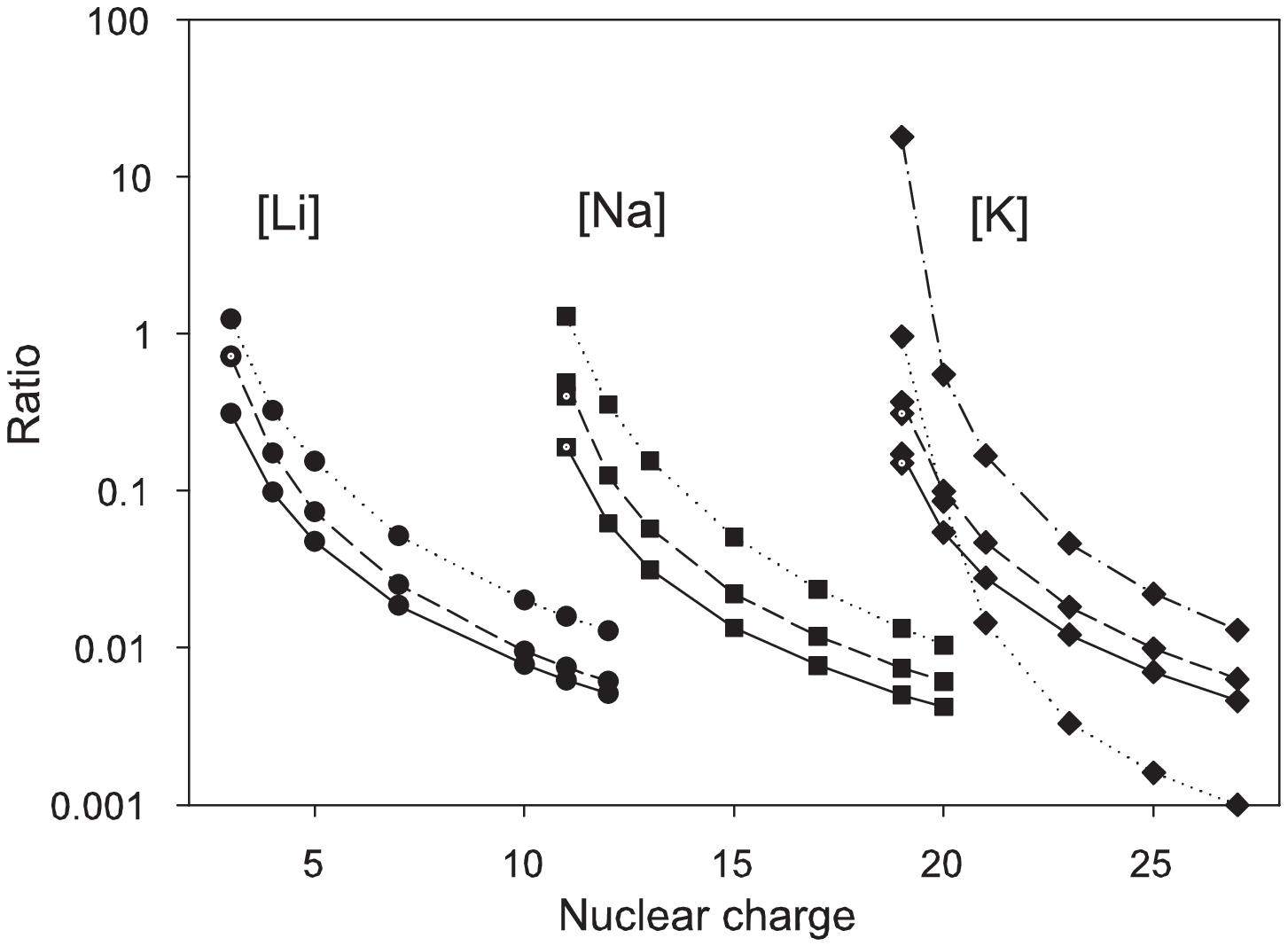}
\end{center}
\end{figure}